\documentclass[prd,superscriptaddress,amsfonts,amssymb,amsmath,showpacs]{revtex4-2}
\usepackage{bm}
\usepackage{amsfonts}
\usepackage{latexsym}
\usepackage[latin1]{inputenc}
\usepackage{graphicx}
\usepackage{amsmath}
\usepackage{palatino}
\usepackage{mathpazo}
\usepackage{textcomp} 
\usepackage{float}
\usepackage{booktabs}
\usepackage{dcolumn}
\usepackage{ragged2e}
\usepackage{hyperref}
\hypersetup{colorlinks,citecolor=green}
\hypersetup{colorlinks=true,linkcolor=blue,filecolor=blue,    urlcolor=magenta}
\usepackage{amsmath}
\usepackage{xcolor}
\usepackage{orcidlink}
\usepackage[caption=false]{subfig}
\usepackage{commath}
\usepackage[autostyle]{csquotes}
\def\jnl@style{\it}
\def\aaref@jnl#1{{\jnl@style#1}}

\def\aaref@jnl#1{{\jnl@style#1}}

\def\aj{\aaref@jnl{AJ}}                   
\def\apj{\aaref@jnl{ApJ}}                 
\def\apjl{\aaref@jnl{ApJ}}                
\def\apjs{\aaref@jnl{ApJS}}               
\def\apss{\aaref@jnl{Ap\&SS}}             
\def\aap{\aaref@jnl{A\&A}}                
\def\aapr{\aaref@jnl{A\&A~Rev.}}          
\def\aaps{\aaref@jnl{A\&AS}}              
\def\mnras{\aaref@jnl{Mon.~Not.~Roy.~Astron.~Soc.}}             
\def\prd{\aaref@jnl{Phys.~Rev.~D}}        
\def\plb{\aaref@jnl{Phys.~Lett.~B}}        
\def\prc{\aaref@jnl{Phys.~Rev.~C}}  
\def\prl{\aaref@jnl{Phys.~Rev.~Lett.}}    
\def\qjras{\aaref@jnl{QJRAS}}             
\def\skytel{\aaref@jnl{S\&T}}             
\def\ssr{\aaref@jnl{Space~Sci.~Rev.}}     
\def\zap{\aaref@jnl{ZAp}}                 
\def\nat{\aaref@jnl{Nature}}              
\def\aplett{\aaref@jnl{Astrophys.~Lett.}} 
\def\apspr{\aaref@jnl{Astrophys.~Space~Phys.~Res.}} 
\def\physrep{\aaref@jnl{Phys.~Rep.}}      
\def\physscr{\aaref@jnl{Phys.~Scr}}       
\def\commat{\aaref@jnl{Comm.~Math.~Phys.}}              
\def\science{\aaref@jnl{Science}}               
\def\cqg{\aaref@jnl{Classical Quant.~Grav.}}            
\def\jpcs{\aaref@jnl{JPCS}}                                     
\def\ijmpd{\aaref@jnl{Int.~J.~Mod.~Phys.~D}}                    
\def\grg{\aaref@jnl{Gen.~Relat.~Gravit.}}               
\def\rpp{\aaref@jnl{Rep.~Prog.~Phys.}}          
\def\npa{\aaref@jnl{Nucl.~Phys.~A}}        
\def\lrr{\aaref@jnl{Living Rev.~Rel.}}                   
\def\jcap{\aaref@jnl{J.~Cosmology Astropart.~Phys.}}    
\def\rmp{\aaref@jnl{Rev.~Mod.~Phys.}}   
\def\epjc{\aaref@jnl{Eur.~Phys.~J.~C}}

\allowdisplaybreaks[1]

\begin{document}

\color{black}       

\title{White dwarfs in regularized 4D Einstein-Gauss-Bonnet gravity}

\author{Juan M. Z. Pretel \orcidlink{0000-0003-0883-3851}}
 \email{juanzarate@cbpf.br}
 \affiliation{
 Centro Brasileiro de Pesquisas F{\'i}sicas, Rua Dr.~Xavier Sigaud, 150 URCA, Rio de Janeiro CEP 22290-180, RJ, Brazil
}

\author{Takol Tangphati \orcidlink{0000-0002-6818-8404}} 
\email{takoltang@gmail.com}
\affiliation{School of Science, Walailak University, Thasala, \\ Nakhon Si Thammarat, 80160, Thailand}
\affiliation{Research Center for Theoretical Simulation and Applied Research in Bioscience and Sensing, Walailak University, Thasala, Nakhon Si Thammarat 80160, Thailand}

\author{\.{I}zzet Sakall{\i} \orcidlink{0000-0001-7827-9476}}
\email{izzet.sakalli@emu.edu.tr}
\affiliation {Physics Department, Eastern Mediterranean University, Famagusta 99628, North Cyprus via Mersin 10, Turkey.}

\author{Ayan Banerjee \orcidlink{0000-0003-3422-8233}} 
\email{ayanbanerjeemath@gmail.com}
\affiliation{Astrophysics and Cosmology Research Unit, School of Mathematics, Statistics and Computer Science, University of KwaZulu--Natal, Private Bag X54001, Durban 4000, South Africa}


\date{\today}


\begin{abstract}
White dwarfs (WDs), as the remnants of low to intermediate-mass stars, provide a unique opportunity to explore the interplay between quantum mechanical degeneracy pressure and gravitational forces under extreme conditions. In this study, we examine the structure and macroscopic properties of WDs within the framework of 4D Einstein-Gauss-Bonnet (4DEGB) gravity, a modified theory that incorporates higher-order curvature corrections through the Gauss-Bonnet coupling constant $\alpha$. Using the modified Tolman-Oppenheimer-Volkoff (TOV) equations tailored for 4DEGB gravity, we analyze the hydrostatic equilibrium of WDs modeled with a realistic equation of state (EoS). Our findings reveal that the inclusion of the Gauss-Bonnet (GB) term significantly influences the mass-radius ($M-R$) relation, allowing for deviations from the Chandrasekhar mass limit. In particular, we observe that such stars become more compact and slightly smaller with the increase of the parameter $\alpha$. For WDs with $\vert\alpha\vert \leq 500\, \rm km^2$, the impact of 4DEGB gravity appears to be negligible. However, a larger range for $\alpha$ allows for appreciable changes in the $M-R$ diagram, mainly in the high-central-density region. Furthermore, we explore the role of anisotropic pressures, quantified by the parameter $\beta$, on such systems and demonstrate their impact on stability and compactness. For sufficiently large values of $\vert\beta\vert$ keeping negative $\beta$ with a large and positive $\alpha$, there exists a second stable branch according to the classical stability criterion $dM/d\rho_c >0$. These results suggest that anisotropic WDs in 4DEGB gravity exhibit unique characteristics that distinguish them from their general relativistic counterparts, offering a novel testing ground for modified gravity theories in astrophysical settings.
\end{abstract}

\maketitle

\section{Introduction}

WDs \cite{Chandrasekhar1931, Chandrasekhar1967, Tremblay2024} represent one of the most well-understood classes of compact stellar remnants, forming as the final evolutionary state of low to intermediate-mass stars. These objects are supported against gravitational collapse primarily by electron degeneracy pressure, as described by quantum mechanics \cite{Koester1990}. The study of WDs has been pivotal in understanding stellar evolution, binary star systems, and supernova mechanisms \cite{Mazzali2007, Istrate2016, Saumon2022, Caiazzo2023}. Despite the empirical success of general relativity (GR) in modeling the macroscopic properties of compact objects, alternative theories of gravity offer a fertile ground to test the robustness of GR under extreme conditions. One such theory, the regularized 4DEGB gravity, extends GR by incorporating higher-order curvature corrections in the gravitational action \cite{Hennigar2020, Fernandes2020,Ovgun2018}.

Modified theories of gravity \cite{Koyama2016, Joyce2016, Nojiri2017, Kase2019, Langlois2019, Shankaranarayanan2022} are motivated by several fundamental challenges in modern physics, including the quantization of gravity, the nature of singularities, and the cosmological constant problem. Among these, higher-curvature theories (HCTs) stand out as they modify the conventional Einstein-Hilbert action by introducing nonlinear terms in the curvature tensor. These modifications are expected to play a significant role in high-energy astrophysical scenarios, including those involving WDs \cite{Nunes2025}.

Einstein-Gauss-Bonnet (EGB) gravity is a widely studied higher-curvature theory \cite{Lovelock1971, Maharaj2015, Ghosh2017, Aspeitia2021, Chatterjee2022, Bhattacharya2023, Bogadi2023, Pretel2024NPB}, originally formulated in dimensions $D > 4$. The Lagrangian in such gravitational theory includes the Gauss-Bonnet (GB) contribution which is quadratic in the Riemann tensor and is consistent with the low-energy effective action in heterotic string theory. The GB term in the action,
\begin{equation}
\mathcal{G} = R^{\mu\nu\rho\tau}R_{\mu\nu\rho\tau} - 4 R^{\mu\nu}R_{\mu\nu} + R^2,
\end{equation}
is a topological invariant in four dimensions, and thus does not contribute to the field equations. However, recent developments have shown that a well-defined $D \to 4$ limit can be derived by rescaling the GB coupling constant $\alpha$ as
\begin{equation}
\lim_{D \to 4} (D-4) \alpha \to \alpha,
\end{equation}
leading to the 4DEGB theory \cite{Hennigar2020, Fernandes2020, Glavan2020}. This theory introduces a scalar field $\phi$ coupled to the GB term, resulting in a scalar-tensor modification of GR. The action for 4DEGB gravity is given by \cite{Gammon2024}:
\begin{equation}\label{ModAction}
S = \frac{1}{2\kappa}\int d^4x \sqrt{-g} \left[ R + \alpha \left( \phi \mathcal{G} + 4 G_{\mu\nu} \nabla^\mu \phi \nabla^\nu \phi - 4 (\nabla \phi)^2 \Box \phi + 2 (\nabla \phi)^4 \right) \right] + S_m,
\end{equation}
where $\kappa= 8\pi$, $R$ is the Ricci scalar and $S_m$ is the matter action. The modification of GB provides a unique framework to explore the impact of higher-curvature corrections on compact objects like quark stars (QSs) \cite{Gammon2024, Doneva2021, Banerjee2021}, electrically charged quark stars \cite{Pretel2022EPJC, Gammon2025arxiv} and neutron stars (NSs) \cite{Doneva2021, Saavedra2024}. As will be seen later in our results, the inclusion of the GB term and its associated coupling constant $\alpha$ introduces new phenomenological features that may alter the mass-radius relation and stability conditions \cite{Banerjee2025,Bora2023,Banerjee2024} of WDs, mainly in the high central density region. For the first time, to the best of our knowledge, this work addresses the study of WDs in 4DEGB gravity under the presence of isotropic and anisotropic pressure.

Observational astrophysics has increasingly unveiled deviations in the properties of compact objects that challenge the predictions of standard GR-based models. Among these are the discoveries of unusually light WDs and other compact remnants, whose physical characteristics strain the theoretical boundaries imposed by conventional EoSs. Such anomalies have sparked considerable interest in exploring alternative frameworks that could accommodate these deviations. For instance, the Chandrasekhar mass limit \cite{Chandrasekhar1931}, which represents the maximum mass a non-rotating WD can achieve before collapsing into a neutron star or black hole, arises naturally within the context of GR due to the balance between gravitational forces and electron degeneracy pressure \cite{Mathew2017, Carvalho2018}. Nevertheless, this critical mass is fundamentally dependent on the underlying gravitational theory \cite{Carvalho2017, Astashenok2022, Kalita2022, Lobato2022, Li2024}, and modifications introduced by alternative gravity frameworks, such as higher-curvature theories, have the potential to shift or redefine this established limit. These modifications could result in more massive or more compact WDs than those allowed under GR, with significant implications for their stability and observable properties. Additionally, deviations from GR may offer explanations for peculiar observational phenomena, such as compact remnants that occupy the so-called mass gap between NSs and black holes or unusually small-radius WDs that defy the constraints imposed by GR. By introducing corrections to the gravitational interaction, alternative theories like 4DEGB could provide a more comprehensive understanding of these anomalies. Investigating how such theories alter the structure and stability of WDs not only helps to explain these unusual observations but also serves as a critical test of GR's validity in extreme environments, offering potential avenues for discovering new physics beyond the standard model of gravity.

In this paper, we investigate the structure and properties of WDs within the framework of 4DEGB gravity, focusing on how the modifications introduced by the GB coupling parameter $\alpha$ affect their fundamental characteristics. WDs, being ideal laboratories for exploring the interplay between quantum mechanical degeneracy pressure and gravitational forces, provide an excellent testbed for assessing the predictions of modified gravity theories. By incorporating the GB corrections, we aim to evaluate their impact on key features such as the mass-radius relation, the maximum mass limit, and the overall stability of these compact stars. Specifically, the theory's deviations from GR may introduce novel astrophysical phenomena that could distinguish it from standard GR predictions. To achieve this, we solve the modified TOV equations for an anisotropic stellar fluid, which govern the hydrostatic equilibrium of spherically symmetric objects in 4DEGB gravity. These equations, derived by incorporating the additional contributions from the GB term and its coupling with the scalar field, require careful numerical treatment to explore a wide parameter space of $\alpha$ values, while the degree of anisotropy is measured by the parameter $\beta$.

Since anisotropic pressure introduces substantial changes in the macroscopic properties of relativistic compact stars \cite{Biswas2019, Pretel2020, Rahmansyah2021, Curi2022, Pretel2024PLB, Becerra2024, Lopes2024, Mohanty2024, Pretel2024}, such as QSs and NSs, then the interest of including a tangential pressure in the present study arises. Coupled with an appropriate EoS for WD matter and a phenomenological anisotropy profile, the solutions of the modified stellar structure equations enable us to analyze how the physical characteristics of (an)isotropic WDs, such as their radii, radial and transverse pressures, and compactness, are modified under this gravitational framework. Such exploration paves the way for a more profound comprehension of the intricate relationship between gravitational forces and high-density matter. In summary, we aim to provide new insights into the astrophysical implications of 4DEGB gravity, offering a compelling case for its consideration as an alternative to GR in modeling compact stellar remnants.

Our work is motivated not only by mathematical curiosity but also by the physical insights offered by 4DEGB gravity. Unlike simpler modified gravity theories such as $f(R)$ or $f(T)$, 4DEGB gravity emerges naturally from string theory, providing a more fundamental basis to explore deviations from the Chandrasekhar mass limit \cite{Shankaranarayanan2022}. In this study, we demonstrate that while standard values of the GB coupling constant $\alpha$ yield minimal changes in WD properties, extending the range of $\alpha$ significantly alters the mass-radius relation. In addition, the inclusion of an anisotropic pressure parameter $\beta$ introduces additional observable characteristics, such as the appearance of two stable branches, which can serve as distinctive signatures to discriminate between competing theories. The rest of this paper is organized as follows: In Section \ref{sec:theory}, we provide an overview of the theoretical framework, including the field equations and the modifications to the TOV equations in 4DEGB gravity. Section \ref{sec:EoS} discusses the equations of state used to model WD matter as well as the anisotropy model. Section \ref{sec:results} presents the numerical results, including the mass-radius relationships and stability analysis for isotropic and anisotropic configurations. Finally, Section \ref{sec:conclusion} concludes with a discussion of the implications of our findings and possible directions for future research.

\section{Theoretical Framework}\label{sec:theory}
\subsection{Field equations in 4DEGB gravity}

WDs, being highly dense stellar remnants, are excellent candidates for testing modifications to GR, particularly in the context of higher-curvature theories like 4DEGB gravity. The incorporation of the GB term into the gravitational action provides a natural extension to GR by introducing corrections that become relevant in strong gravitational regimes. These corrections are encoded in the GB coupling constant $\alpha$, which quantifies the contribution of the higher-order curvature terms to the dynamics of spacetime.

Considering the action of 4DEGB gravity \eqref{ModAction}, which combines the Einstein-Hilbert term with a regularized GB term, the variation with respect to the scalar field $\phi$ leads to its equation of motion:
\begin{equation}\label{eq:eomscalar}
\begin{aligned}
\mathcal{E}_{\phi} = &-\mathcal{G} + 8 G^{\mu \nu} \nabla_{\nu} \nabla_{\mu} \phi + 8 R^{\mu \nu} \nabla_{\mu} \phi \nabla_{\nu} \phi - 8(\Box \phi)^2 + 8(\nabla \phi)^2 \Box \phi + 16 \nabla^{\mu} \phi \nabla^{\nu} \phi \nabla_{\nu} \nabla_{\mu} \phi\\
& + 8 \nabla_{\nu} \nabla_{\mu} \phi \nabla^{\nu} \nabla^{\mu} \phi = 0.
\end{aligned}
\end{equation}
The variation with respect to the metric $g_{\mu\nu}$ yields the modified Einstein field equations:
\begin{equation}\label{eq:eommetric}
\begin{aligned}
\mathcal{E}_{\mu \nu} =&\ G_{\mu \nu}+\alpha\left\lbrace\phi H_{\mu \nu}-2 R\left[\left(\nabla_{\mu} \phi\right)\left(\nabla_{\nu} \phi\right)+\nabla_{\nu} \nabla_{\mu} \phi\right]+8 R_{(\mu}^{\sigma} \nabla_{\nu)} \nabla_{\sigma} \phi+8 R_{(\mu}^{\sigma}\left(\nabla_{\nu)} \phi\right)\left(\nabla_{\sigma} \phi\right)\right.\\
&-2 G_{\mu \nu}\left[(\nabla \phi)^{2}+2 \square \phi\right]-4\left[\left(\nabla_{\mu} \phi\right)\left(\nabla_{\nu} \phi\right)+\nabla_{\nu} \nabla_{\mu} \phi\right] \square \phi-\left[g_{\mu \nu}(\nabla \phi)^{2}-4\left(\nabla_{\mu} \phi\right)\left(\nabla_{\nu} \phi\right)\right](\nabla \phi)^{2} \\
&+8\left(\nabla_{(\mu} \phi\right)\left(\nabla_{\nu)} \nabla_{\sigma} \phi\right) \nabla^{\sigma} \phi-4 g_{\mu \nu} R^{\sigma \rho}\left[\nabla_{\sigma} \nabla_{\rho} \phi+\left(\nabla_{\sigma} \phi\right)\left(\nabla_{\rho} \phi\right)\right]+2 g_{\mu \nu}(\square \phi)^{2} \\
& -4 g_{\mu \nu}\left(\nabla^{\sigma} \phi\right)\left(\nabla^{\rho} \phi\right)\left(\nabla_{\sigma} \nabla_{\rho} \phi\right)+4\left(\nabla_{\sigma} \nabla_{\nu} \phi\right)\left(\nabla^{\sigma} \nabla_{\mu} \phi\right) \\ 
&\left. -2 g_{\mu \nu}\left(\nabla_{\sigma} \nabla_{\rho} \phi\right)\left(\nabla^{\sigma} \nabla^{\rho} \phi\right)
+4 R_{\mu \nu \sigma \rho}\left[\left(\nabla^{\sigma} \phi\right)\left(\nabla^{\rho} \phi\right)+\nabla^{\rho} \nabla^{\sigma} \phi\right] \right\rbrace = \kappa T_{\mu \nu},
\end{aligned}
\end{equation}
where $T_{\mu \nu}$ is the energy-momentum tensor of the matter content, defined by the usual form
\begin{equation}
    T_{\mu\nu} = \frac{-2}{\sqrt{-g}}\frac{\delta S_m}{\delta g^{\mu\nu}} .
\end{equation}

Additionally, $H_{\mu \nu}$ is the GB tensor, which is given by
\begin{equation}\label{eq:gbtensor}
H_{\mu \nu} = 2 \left[ R R_{\mu \nu} - 2 R_{\mu \alpha \nu \beta} R^{\alpha \beta} + R_{\mu \alpha \beta \sigma} R_{\nu}^{\alpha \beta \sigma} - 2 R_{\mu \alpha} R_{\nu}^{\alpha} - \frac{1}{4} g_{\mu \nu} \mathcal{G} \right].
\end{equation}
The above field equations satisfy the following relationship
\begin{equation}\label{eq:fieldeqntrace}
\kappa g^{\mu \nu}T_{\mu \nu} =g^{\mu \nu} \mathcal{E}_{\mu \nu}+\frac{\alpha}{2} \mathcal{E}_{\phi}= -R-\frac{\alpha}{2} \mathcal{G},
\end{equation}
and this can act as a useful consistency check to see whether prior solutions generated via the Glavin/Lin method are even possible solutions to the gravity theory. The resulting equations form a scalar-tensor theory of gravity where the scalar field $\phi$ interacts with the spacetime curvature. These equations retain the second-order nature of GR while allowing for new phenomenological predictions in astrophysical contexts.

\subsection{Equilibrium configurations via modified TOV equations}

The modified version of the TOV equations in the context of 4DEGB gravity are derived from the field equations, incorporating the additional contributions from the GB term. To model static, spherically symmetric compact stars, we adopt the usual line element
\begin{eqnarray}\label{metric}
  ds^2 = - e^{2\Phi(r)}dt^2 + e^{2\Psi(r)}dr^2 + r^2d\Omega^2 ,
\end{eqnarray}
where $\Phi(r)$ and $\Psi(r)$ are unknown metric functions to be determined. Furthermore, the anisotropic matter-energy distribution is described by 
\begin{eqnarray}\label{EMT}
T_{\mu \nu} = (\rho + p_{\perp})u_\mu u_\nu + p_{\perp}g_{\mu \nu} - \sigma \chi_{\mu} \chi_{\nu}, 
\end{eqnarray}
where $\rho$ is the energy density, $p_r$ the radial pressure and $p_{\perp}$ is the transverse pressure. Besides, $\sigma\equiv p_{\perp}- p_{r}$ is the anisotropic factor in the stellar source. Consequently, the 00 and 11 components of the field equations generate
\begin{align}
    \frac{2}{r}\left[ 1+ \frac{2\alpha(1-e^{-2\Psi})}{r^2} \right]\frac{d\Psi}{dr} &= e^{2\Psi} \left[ 8\pi\rho - \frac{1- e^{-2\Psi}}{r^2}\left( 1- \frac{\alpha(1- e^{-2\Psi})}{r^2} \right) \right] , \label{FEq1}  \\
    \frac{2}{r}\left[ 1+ \frac{2\alpha(1-e^{-2\Psi})}{r^2} \right]\frac{d\Phi}{dr} &= e^{2\Psi} \left[ 8\pi p_r + \frac{1- e^{-2\Psi}}{r^2}\left( 1- \frac{\alpha(1- e^{-2\Psi})}{r^2} \right) \right] . \label{FEq2}
\end{align}

In addition, the covariant conservation of the energy-momentum tensor provides
\begin{equation}\label{ConsevEq}
    \frac{dp_r}{dr} = -(\rho +p_r)\frac{d\Phi}{dr} + \frac{2}{r}\sigma .
\end{equation}
The relation between the mass function $m(r)$ and metric $\Psi(r)$ is given by the usual way \cite{Hennigar2020, Gammon2024}
\begin{equation}\label{PsiEq}
    e^{-2\Psi} = 1+ \frac{r^2}{2\alpha}\left[ 1- \sqrt{1+ \frac{8\alpha m}{r^3}} \right] ,
\end{equation}
namely, $m(r)$ is the enclosed gravitational mass within the radial coordinate $r$. Note also that in the limit $\alpha \rightarrow 0$, the last expression behaves asymptotically as
\begin{equation}
    \lim_{\alpha \rightarrow 0} e^{-2\Psi} = 1- \frac{2m}{r} + \frac{4m^2}{r^4}\alpha + \cdots .
\end{equation}

In view of Eqs.~\eqref{FEq2} and \eqref{PsiEq}, we obtain 
\begin{equation}
    \frac{d\Phi}{dr} = \frac{r^3 \left(\sqrt{1+ \frac{8 \alpha  m}{r^3}}+8 \pi  \alpha  p_r-1\right)-2 \alpha  m}{r^2
   \left(2 \alpha +r^2\right) \sqrt{1+ \frac{8 \alpha  m}{r^3}}-8 \alpha  m r-r^4} ,
\end{equation}
so that Eqs.~\eqref{FEq1} and \eqref{ConsevEq} become respectively
\begin{align}
    \frac{dm}{dr} &= 4\pi r^2\rho ,  \label{TOV1}  \\
    \frac{dp_r}{dr} &= \frac{(\rho+ p_r)\left[ 2\alpha m + r^3 (1- \mathcal{A}- 8\pi\alpha p_r) \right]}{r^2\mathcal{A}\left( r^2+ 2\alpha - r^2\mathcal{A} \right)} + \frac{2}{r}\sigma ,  \label{TOV2} 
\end{align}
where we have defined $\mathcal{A} \equiv \sqrt{1+ \frac{8\alpha m}{r^3}}$ .

The above differential equations \eqref{TOV1} and \eqref{TOV2} are known as modified TOV equations in 4DEGB gravity and reduce to the isotropic case when $\sigma= 0$ \cite{Gammon2024}. To close the system of equations, an EoS that relates $p_r$ and $\rho$ must be supplied. We will also adopt an anisotropy profile that relates $\sigma$ to radial pressure and mass function, so that we have only two unknown variables (i.e., $\rho$ and $m$) to determine. Therefore, we will employ both the Chandrasekhar EoS for WD matter and the QL model to account for anisotropic pressures. We will discuss this in more detail below.

\subsection{Boundary and initial conditions}

As in conventional Einstein gravity, the boundary conditions for the numerical integration are specified at the stellar center ($r = 0$) and the surface ($r = R$). At the center, the enclosed mass is zero, $m(0) = 0$, and the central mass density $\rho(0)= \rho_c$ is provided as an input parameter. The integration proceeds outward until the radial pressure vanishes, $p_r(R) = 0$, which defines the stellar radius $R$. The mass at this radius corresponds to the total mass of the compact star, i.e.,~$M = m(R)$. The variation of the input parameter $\rho_c$ will allow us to obtain a family of anisotropic WDs represented in an $M-R$ diagram.

For anisotropic configurations, additional conditions are imposed on the different physical quantities in the specific anisotropy model, ensuring that the radial pressure is equal to the tangential pressure at the center of the fluid sphere. Other conditions are also required, such as gradients for mass density and radial pressure must be negative, the radial and tangential speed of sound should be less than the speed of light, etc.~\cite{Mak2003}. As we will see below, this model satisfies such physical acceptability conditions. It is worth emphasizing that the radial speed of sound inside WDs is always smaller than the speed of light, and since it depends exclusively on the EoS, then this causality condition is trivially satisfied in our study.

\section{E\lowercase{o}S for WD\lowercase{s} and anisotropy model}\label{sec:EoS}

The EoS plays a crucial role in modeling the equilibrium structure of WDs by providing a functional relation between radial pressure and energy density. Such relation is essential for solving the modified TOV equations within the framework of 4DEGB gravity. WDs are primarily supported by electron degeneracy pressure, which can be effectively described using the Chandrasekhar EoS \cite{Chandrasekhar1935}. Additionally, to capture potential anisotropic effects in stellar matter, we employ the Quasi-Local (QL) model \cite{Horvat2011}, which introduces a tangential pressure in addition to the radial pressure into the stellar system. It was shown that the main impact of GR with respect to the Newtonian context occurs on WDs with masses greater than $1.3\, M_\odot$, indicating that general relativistic effects become important for massive WDs \cite{Carvalho2018}. This has motivated the study of WDs in modified gravity theories, such as $R$-squared gravity \cite{Astashenok2022, Kalita2022}, linear $f(R,T)$ gravity \cite{Carvalho2017}, $f(R,L_m)$ theories \cite{Lobato2022} and Rastall-Rainbow gravity \cite{Li2024}. Furthermore, in standard Einstein gravity, several authors have shown that anisotropic pressure induces significant changes in the global properties of massive compact stars \cite{Biswas2019, Pretel2020, Rahmansyah2021, Curi2022, Pretel2024PLB, Becerra2024, Lopes2024, Mohanty2024, Pretel2024}. These anisotropy effects are therefore expected to be noticeable in massive WDs. In that regard, in the present study we will also consider tangential pressure in WDs within the framework of 4DEGB gravity.

\subsection{Chandrasekhar EoS}

The Chandrasekhar EoS \cite{Chandrasekhar1935}, a cornerstone of WD modeling, describes the pressure-density relationship for a relativistic degenerate electron gas. The radial pressure $p_r$ as a function of the Fermi momentum $k_F$ is given by
\begin{equation}
p_r(k_F) = \frac{1}{3\pi^2 \hbar^3} \int_0^{k_F} \frac{k^4}{\sqrt{k^2 + m_e^2}} \, dk,
\end{equation}
where $\hbar$ is the reduced Planck constant, $m_e$ is the electron mass, and $k$ represents the momentum of the electrons. This integral can be evaluated to yield
\begin{equation}\label{EoSpart1}
p_r(k_F) = \frac{\pi m_e^4}{3 h^3} \left[ x_F (2x_F^2 - 3)\sqrt{x_F^2 + 1} + 3 \sinh^{-1}x_F \right],
\end{equation}
with $h = 2\pi\hbar$ being the Planck constant, and $x_F = p_F / m_e c$ is the dimensionless Fermi momentum. The corresponding mass density $\rho$ is expressed as
\begin{equation}\label{EoSpart2}
\rho = \frac{8 \pi \mu_e m_H m_e^3}{3 h^3} x_F^3,
\end{equation}
where $\mu_e$ is the mean molecular weight per electron (we adopt $\mu_e = 2$ for our analysis) and $m_H$ is the mass of a hydrogen atom. 

This EoS describes the essential physics of degenerate matter in WD stars, where pressure arises from the quantum mechanical effects of Pauli exclusion rather than thermal motion. In the high-density regime, the relativistic corrections included in the Chandrasekhar EoS become crucial for accurately predicting the mass-radius diagram and stability of WDs. \textcolor{black}{This model based on an ideal Fermi gas at zero temperature was recently used to examine the properties of anisotropic white dwarf stars within the framework of Rainbow gravity \cite{Tangphati2025}.}

\subsection{Incorporating anisotropic pressure with the QL model}

While isotropic pressure distributions are traditionally assumed in WD modeling, there are several physical mechanisms that could potentially introduce anisotropy in these systems. As a matter of fact, the fluid pressures become anisotropic in the presence of strong magnetic fields \cite{Sinha2013, Paulucci2011}. In that regard, WDs as polytropes for anisotropic fluids were considered to estimate the maximum stable mass of magnetized WDs, which could be greater than $3\, M_\odot$ \cite{Das2014}. For further studies on the relationship between strong magnetic fields and anisotropic pressure in WDs, we refer the reader to Refs.~\cite{Das2012PRD, Deb2022ApJ}. Accordingly, strong magnetic fields (which are commonly observed in WDs \cite{Terada2008, Reimers1996}) can create pressure anisotropies by affecting the motion of charged particles differently along and perpendicular to field lines. Similarly, rapid rotation can induce pressure differentials between different directions. For instance, strong magnetic fields (observed up to $10^9G$ in some WDs), rapid rotation, or even crystallization effects at high densities can create directional pressure differences. In this work, we adopt the QL anisotropy model as a phenomenological framework to capture these potential effects in 4DEGB gravity \cite{Shankaranarayanan2022,Pretel2020}. While the exact physical origins of anisotropy in WDs require further study, our approach offers a versatile starting point to explore its impact on stellar stability and structure.

To extend the analysis to cases where the pressure may not be isotropic, we incorporate the QL model proposed by Horvat \textit{et al.}~\cite{Horvat2011}. Anisotropic pressure can arise in highly dense systems where the radial pressure $p_r$ differs from the tangential pressure $p_\perp$. The degree of anisotropy is characterized by the parameter $\sigma$, defined as follows
\begin{equation}\label{AnisoModelEq}
\sigma \equiv p_\perp - p_r = \beta p_r \mu,
\end{equation}
where $\beta$ is a constant that quantifies the extent of anisotropy, and $\mu = \frac{2m(r)}{r}$ represents the local compactness of the star. The free parameter $\beta$ is limited within the range $[-2, 2]$, see for example Refs.~\cite{Pretel2020, Rahmansyah2021, Curi2022, Pretel2024PLB, Becerra2024, Mohanty2024, Silva2014, Pretel2022CQG, Pretel2022MPLA, Tangphati2023, Becerra2024b} for typical values. A positive $\beta$ corresponds to stellar systems where $p_\perp > p_r$, while a negative $\beta$ indicates the opposite. 

The choice of anisotropic pressure becomes particularly significant in dense regions of the star, where deviations from the isotropic context could influence the star's stability and maximum mass. In the QL model, the anisotropy naturally vanishes as $r \to 0$, recovering the isotropic limit at the stellar center. Additionally, the model ensures that the radial and tangential pressures smoothly vanish at the stellar surface, satisfying the physical boundary conditions
\begin{equation}
p_r(r \to R) = 0,  \qquad  p_\perp(r \to R) = 0.
\end{equation}

We acknowledge that the specific physical origin of anisotropy in WDs requires further detailed investigation. Nonetheless, by including this parameter in our analysis, we provide a more comprehensive framework for future studies that might incorporate specific physical mechanisms for anisotropy in WDs. Our study employs the QL anisotropy model as a phenomenological approach to explore how pressure anisotropies might affect WD structure in 4DEGB gravity.

\subsection{Physical implications of EoS and anisotropy}

The combined use of the Chandrasekhar EoS and the quasi-local ansatz provides a comprehensive framework for modeling WDs. The Chandrasekhar EoS captures the quantum mechanical effects governing electron degeneracy pressure, while the QL anisotropy model accounts for anisotropic pressures that may arise due to strong gravitational fields or interactions in dense stellar matter. As we will see later, the inclusion of anisotropy is particularly relevant in the context of 4DEGB gravity, where higher-curvature corrections could enhance or suppress deviations from the isotropic case.

In subsequent sections, we will employ the above EoS along with Eq.~\eqref{AnisoModelEq} to solve the modified TOV equations numerically and investigate the impact of 4DEGB corrections and tangential pressure on the $M-R$ relation and compactness ($C$) of WDs.

\section{Numerical implementation and computational setup} \label{sec:results}

The GB coupling constant $\alpha$ introduces a new degree of freedom in the gravitational dynamics, and here we will numerically examine its effects on the relativistic structure of WDs. To investigate such as equilibrium configurations in the framework of 4DEGB gravity, we numerically solve the modified TOV equations \eqref{TOV1}-\eqref{TOV2}. This involves employing appropriate boundary conditions, EoS, and a robust numerical scheme to explore the effects of the GB coupling parameter $\alpha$ on the mass-radius diagram and stability properties of WDs. The parameter space explored includes both isotropic and anisotropic configurations of the stellar structure.

\subsection{Numerical approach}

The stellar structure equations \eqref{TOV1}-\eqref{TOV2} are solved using a fourth-order Runge-Kutta integration method. The equations are discretized on a fine grid, and care is taken to ensure numerical stability and convergence. A parametric study is initially conducted for a given value of central density $\rho_c$ and GB coupling constant $\alpha$, spanning values in the range $\alpha \in [-1.0, 1.0]\times 10^4\, \rm{km^2}$ by considering the isotropic case (i.e., when $\beta= 0$). Then the integration will be done over a wide range of central densities for both isotropic and anisotropic configurations.

To visualize our numerical results, $M-R$ diagrams and mass-central density ($M-\rho_c$) relations are constructed for various values of $\alpha$ and $\beta$. These diagrams provide insight into the dependence of stellar properties on $\alpha$, the EoS and anisotropy profile. Both isotropic and anisotropic configurations are analyzed, highlighting the role of pressure anisotropy in modifying the compactness and stability of WDs.

\begin{figure}
    \centering
\includegraphics[width=176mm,scale=0.4]{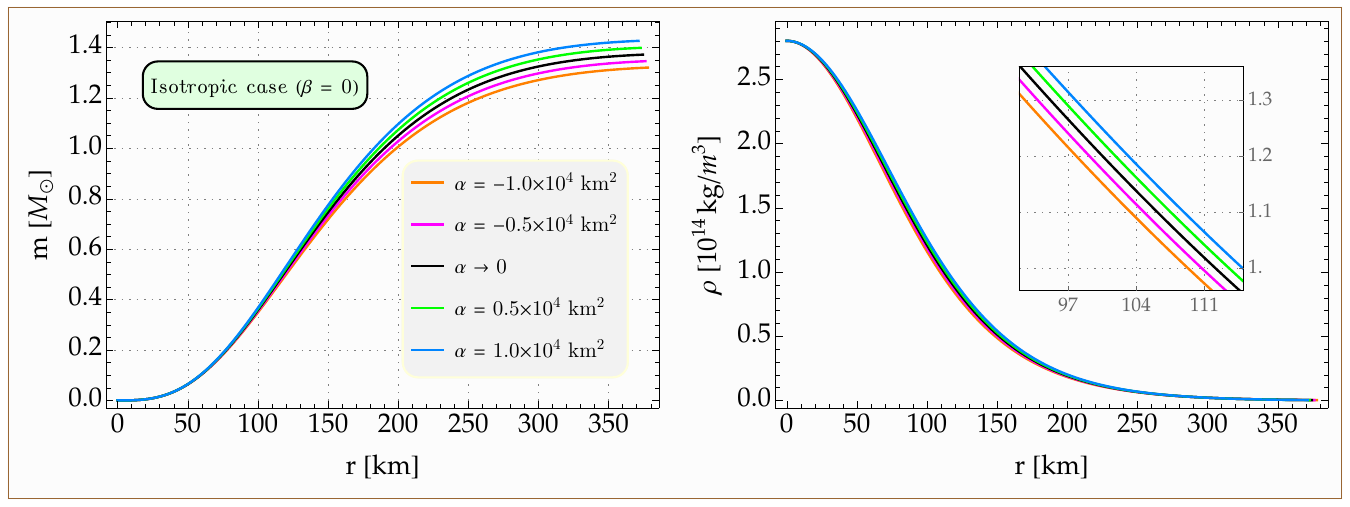}
\includegraphics[width=176mm,scale=0.4]{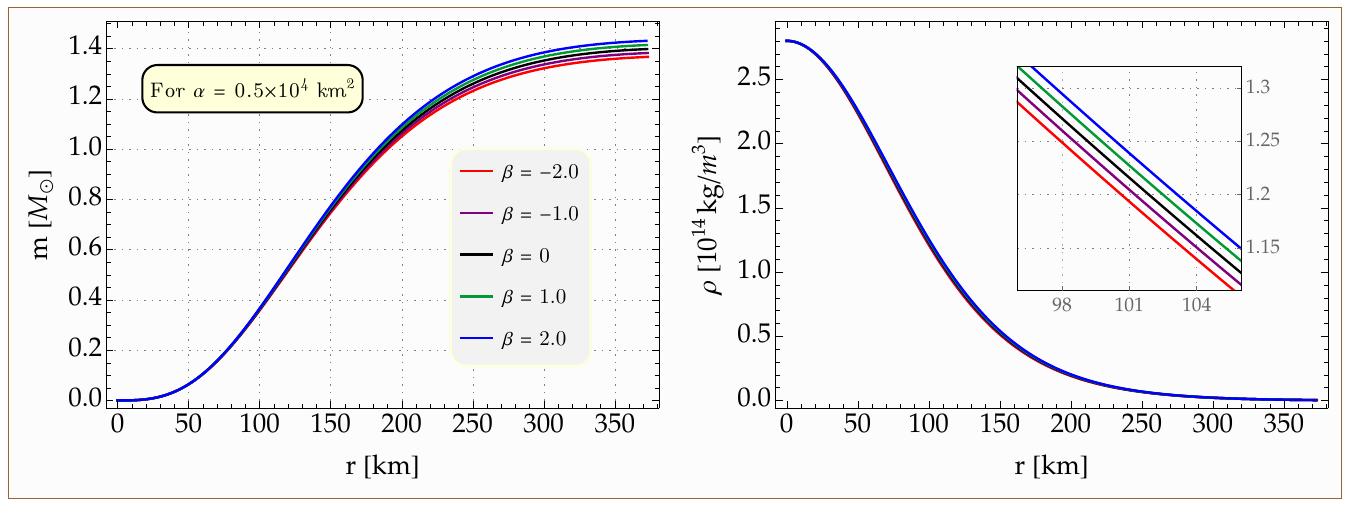}
    \caption{Radial behavior of the mass function (left plots) and energy density (right plots) inside a WD with central density $\rho_c = 2.8 \times 10^{14}\, \rm kg/m^3$ in 4DEGB gravity. The upper panel corresponds to isotropic solutions (i.e., when $\beta =0$) for various values of $\alpha$, while the lower panel represents anisotropic solutions for a fixed $\alpha= 5000\, \rm km^2$. }
    \label{FigRadBeh}
\end{figure}

\subsection{Effects of GB coupling constant}

Given a central density value $\rho_c = 2.8 \times 10^{14}\, \rm kg/m^3$, we begin our analysis by showing the numerical solution of the modified TOV equations for the isotropic case, where $\beta$ is null, see the top panel of Fig.~\ref{FigRadBeh}. The mass $m(r)$ is an increasing function, while the density $\rho(r)$ decreases as we go from the center to the surface of the WD. We observe that positive (negative) values of the GB coupling constant increase (decrease) the mass distribution of the WD, mainly in the outermost regions of the star. Furthermore, the radius of the star $R$ decreases with increasing $\alpha$. Although the effect of $\alpha$ on the mass function is appreciable, the mass density is slightly modified by $\alpha$.

\begin{figure}
    \centering
\includegraphics[width=175mm,scale=0.4]{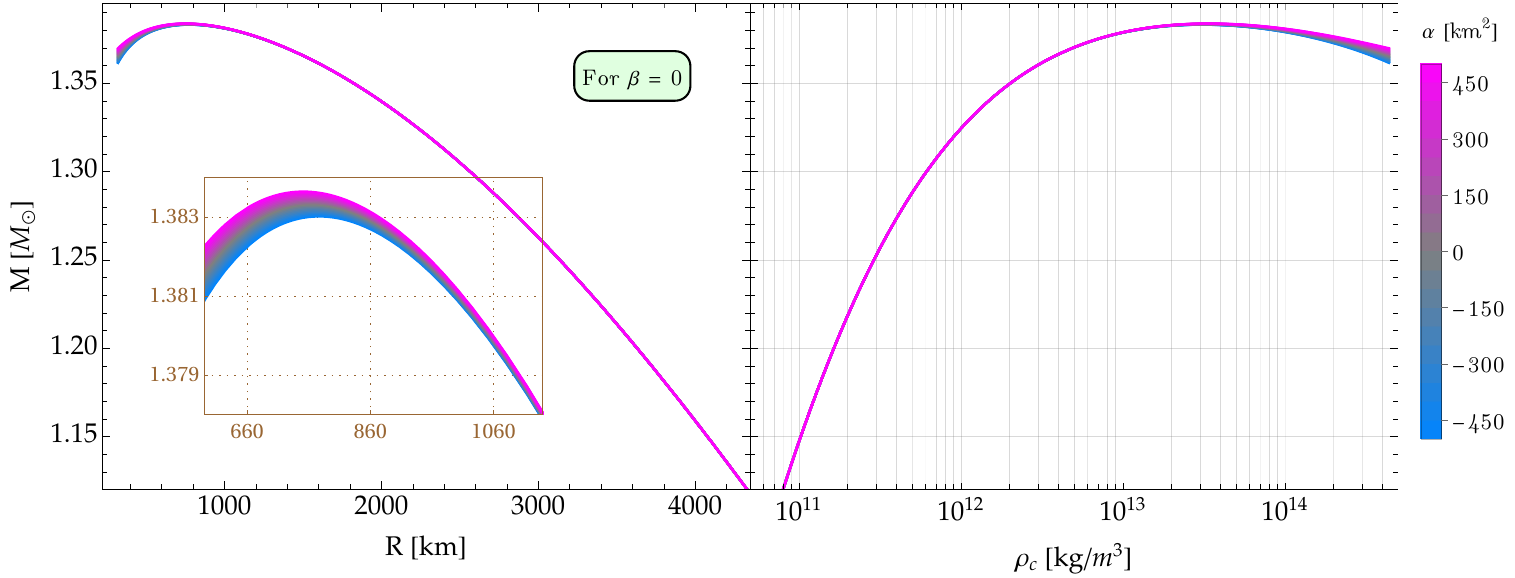}
    \caption{Mass-radius relation (left panel) and mass-central density relation (right panel) for isotropic WDs in 4DEGB gravity, where we have varied the GB coupling parameter in the range $\alpha \in [-500, 500]\, \rm km^2$, see the color scale on the right. The small plot embedded in the left panel is an enlargement of the $M-R$ relation near the maximum-mass points. }
    \label{FigMRRhoSmallAlpha}
\end{figure}

Now we will vary the central density $\rho_c$ to generate a family of equilibrium WDs in 4DEGB gravity considering $\beta= 0$, but before that it is necessary to comment on the values of $\alpha$. Very recently, Saavedra and collaborators \cite{Saavedra2024} have shown that very significant changes in the mass-radius diagrams of NSs due to the GB term occur if $\alpha$ is of the order of $\sim 300\, \rm km^2$, where it is possible to obtain maximum NS masses of up to $\sim 12\, M_\odot$. Taking this work as reference, in Fig.~\ref{FigMRRhoSmallAlpha} we show our numerical results for the mass-radius ($M-R$) relationship (left panel) and the mass-central density ($M-\rho_c$) relation (right panel) for isotropic WDs by using $\alpha \in [-500, 500]\, \rm km^2$, where we have varied the central density up to where the Chandrasekhar EoS is valid, i.e., the neutron drip density $\rho_{\rm drip} \approx 4.3\times 10^{11}\, \rm g/cm^3$ \cite{Shapiro2004}. Above this density value, we would be describing NS interiors and a different EoS would be required than the one established in Eqs.~\eqref{EoSpart1}-\eqref{EoSpart2}. Although for NSs this range of $\alpha$ values produces substantial effects on the radius and mass, we observe that the impact of the GB term is irrelevant for the case of WDs in 4DEGB gravity. Specifically, for WDs with radii greater than $1000\, \rm km$, the impact of 4DEGB gravity appears to be negligible. Slight changes are more noticeable after the maximum mass, from where the WDs are unstable.

\begin{figure}
    \centering
\includegraphics[width=175mm,scale=0.4]{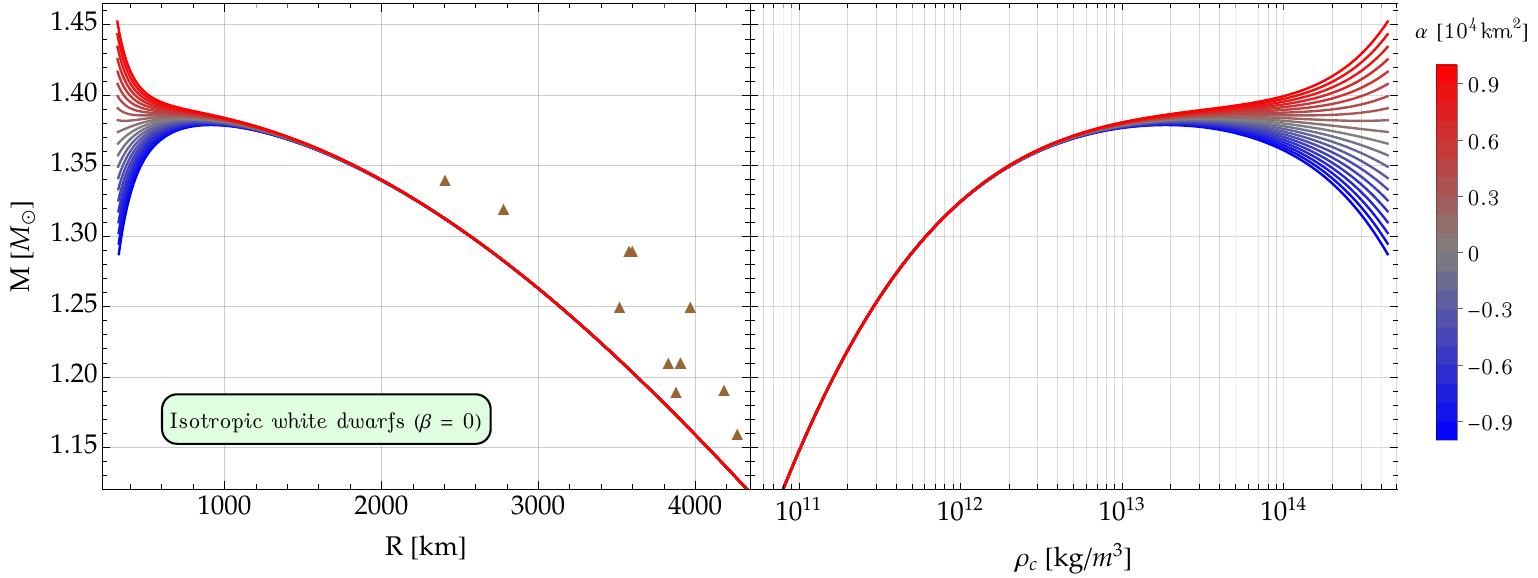}
    \caption{$M-R$ diagram (left panel) and $M-\rho_c$ relation (right panel) for isotropic WDs when the GB parameter varies in the interval $\vert\alpha\vert \leq 1.0\times 10^4\, \rm km^2$. For this range in $\alpha$, the gravitational mass of a WD undergoes appreciable changes for high central densities, namely, for $\rho_c \gtrsim 2.0\times 10^{13}\, \rm kg/m^3$, while below this density value the modifications are irrelevant. \textcolor{black}{The brown triangles on the left plot represent the observational data taken from the catalogue of isolated massive WDs \cite{Nalezyty2004}.} }
    \label{FigMRRho}
\end{figure}

Since the range of values for $\alpha$ in Fig.~\ref{FigMRRhoSmallAlpha} does not generate appreciable changes in the $M-R$ diagram of WDs, we will now use a larger interval, taking advantage of the fact that the observational constraints on the GB coupling parameter have led to $-10^{-36}\, {\rm km^2} < \alpha < 10^4\, \rm km^2$ \cite{Clifton2020, Fernandes2022}. Although the literature also provides lower bounds \cite{Charmousis2022}, it is well known that these constraints depend heavily on the specific EoS used to describe dense stellar matter. Since our study deals with densities lower than those corresponding to NSs or QSs, we can adopt a larger range for $\vert\alpha\vert$. In other words, since we are considering a different range of energy densities (and therefore different stellar systems) than the one already examined in the literature, the GB parameter does not necessarily have to be the same here to appreciate substantial effects on the mass-radius relation due to 4DEGB gravity. Figure \ref{FigMRRho} illustrates our results when the GB coupling parameter $\alpha$ is varied in the range $\alpha \in [-1.0, 1.0]\times 10^4\, \mathrm{km^2}$. Positive values of $\alpha$ generally result in more massive compact configurations (see red curves), while negative $\alpha$ decreases the gravitational mass of these stars (see blue curves). It should be noted that below a central density $\sim 2.0\times 10^{13}\, \rm kg/m^3$, the parameter $\alpha$ has negligible effects; however, the GB coupling constant plays an important role above this central density value.

\textcolor{black}{Unlike WDs in $f(R,T)= R+ 2\lambda T$ gravity where the radius undergoes significant modifications due to the parameter $\lambda$ \cite{Carvalho2017}, in the present study we show that for small masses (or low central densities) the $M-R$ curve is unaltered by the GB term, but the opposite occurs at high central densities. For comparison reasons, in Fig.~\ref{FigMRRho} we have incorporated the observational data taken from the catalogue of isolated massive WDs \cite{Nalezyty2004}. Of course, measurements for several WDs within 100 pc from the Sloan Digital Sky Survey (SDSS) \cite{Kilic2020, Crumpler2024} lie in an even smaller range for the central density where the radii are $\gtrsim 5000\, \rm km$. However, the novelty of our findings lies in the high-mass region, beyond the usual Chandrasekhar limit. Remarkably, our results reveal that for sufficiently large positive $\alpha$ (i.e., $\alpha \gtrsim 2000\, \rm km^2$), it is not possible to obtain a critical WD since the maximum mass cannot be found. Meanwhile, the critical point corresponding to the maximum-mass configuration can be found for negative $\alpha$ (blue curves). This is an unprecedented result for WDs in 4DEGB gravity, something not manifested in GR.} We will comment further on this matter in subsection \ref{SubsecStability}.

\begin{figure*}
    \centering
\includegraphics[width = 8.0cm]{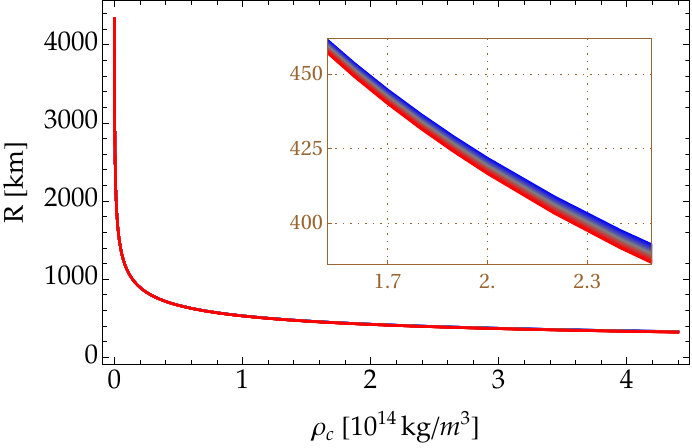}
\includegraphics[width = 9.5cm]{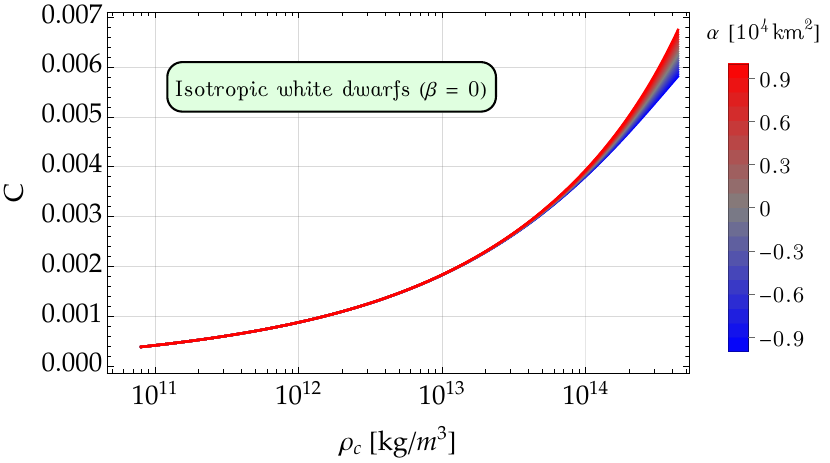}
    \caption{Radius (left) and compactness (right) as a function of central density for the isotropic WDs shown in Fig.~\ref{FigMRRho}. }
    \label{FigRCRho}
\end{figure*}

According to the plot on the left of Fig.~\ref{FigRCRho}, the compact star becomes slightly smaller with increasing $\alpha$, although its mass increases significantly for positive $\alpha$ (i.e., for the smallest stars in the left panel of Fig.~\ref{FigMRRho}). As a consequence of these results, the compactness (given by $C= M/R$) becomes larger with increasing GB coupling parameter. This can be clearly observed in the right plot of Fig.~\ref{FigRCRho}, where $C$ is of the order of $\sim 10^{-3}$. On the other hand, the lower $\alpha$, the less compact are the isotropic WDs in 4DEGB gravity.

The plot on the left of Fig.~\ref{FigCompactness} exhibits the compactness as a function of gravitational mass for the equilibrium configurations with $\beta= 0$. As expected, $C$ changes noticeably due to the GB term only for high central densities, i.e.~for $\rho_c \gtrsim 10^{14}\, \rm kg/m^3$. Specifically, we see that the highest compactnesses are obtained for positive $\alpha$, which is consistent with the right plot of Fig.~\ref{FigRCRho}.

In summary, the inclusion of the GB term significantly affects the maximum mass of WDs (although the radius barely changes) if the range for the coupling constant $\alpha$ is large enough, potentially allowing them to exceed the Chandrasekhar limit. This phenomenon has important implications for understanding the observed properties of unusual compact objects, such as super-Chandrasekhar WDs or low-mass remnants. Nevertheless, when $\alpha$ is of the order of $500\, \rm km^2$, the effects of the GB term on the most basic properties of a WD are negligible (as already illustrated in Fig.~\ref{FigMRRhoSmallAlpha}).

\subsection{Impact of anisotropic pressure}

To examine the effects of anisotropy on WDs in 4DEGB gravity, we solve the modified stellar structure equations for a given central density $\rho_c = 2.8 \times 10^{14}\, \rm kg/m^3$ while holding fixed $\alpha= 5000\, \rm km^2$ as in the bottom panel of Fig.~\ref{FigRadBeh}. For the range $\beta\in [-2.0, 2.0]$, we see that the gravitational mass increases as the anisotropies grow (i.e., for positive values of $\beta$), while the opposite occurs for negative anisotropies. Nonetheless, the changes introduced by $\beta$ on the mass density are negligible with respect to the isotropic case.

\begin{figure}
    \centering
    \includegraphics[width = 17.6cm]{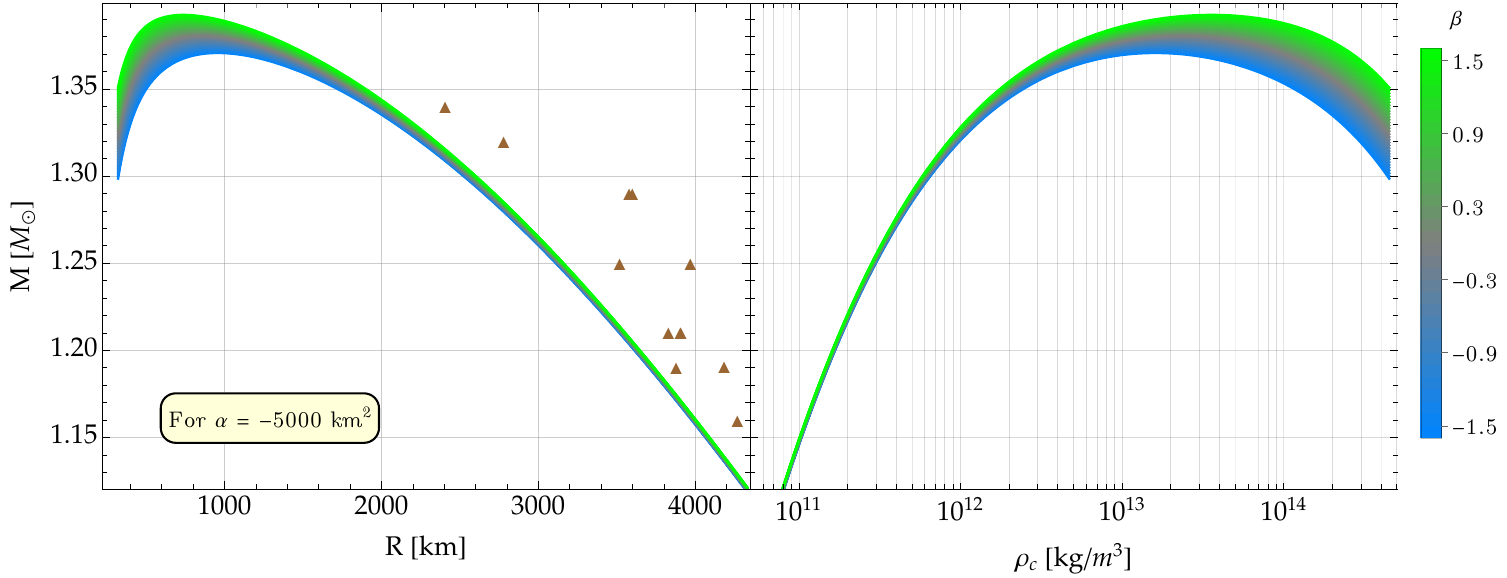}
    \includegraphics[width = 17.6cm]{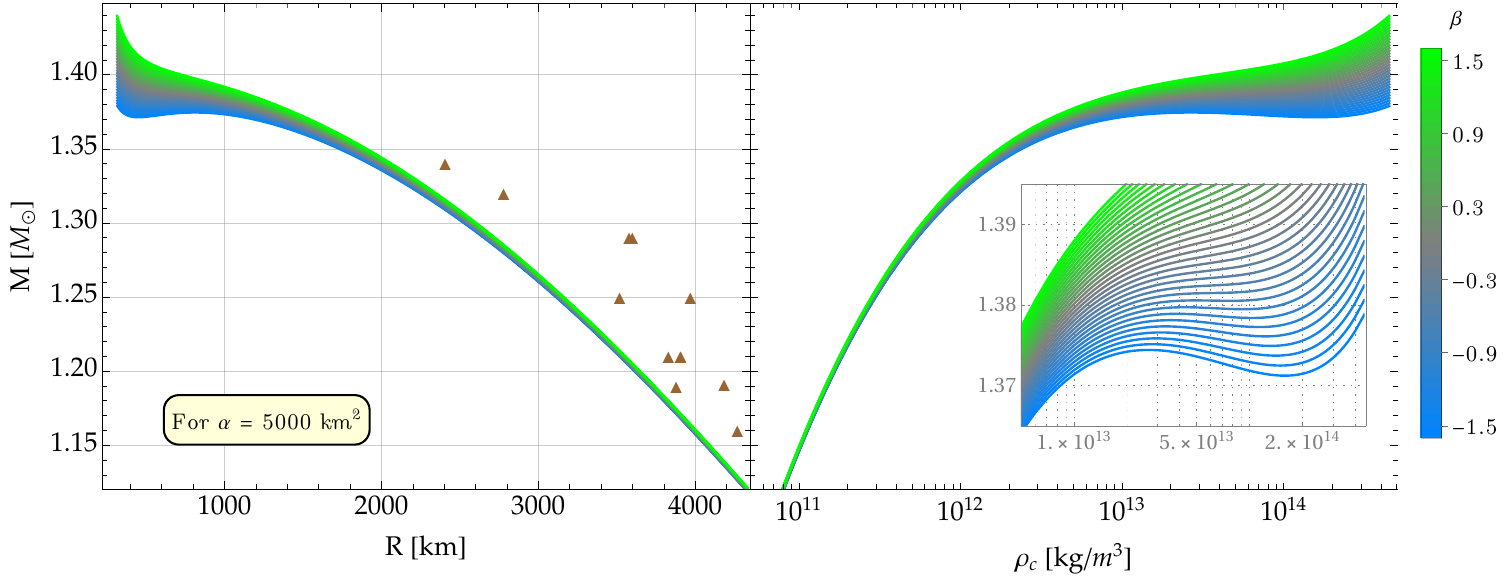}
    \caption{Anisotropic WDs in 4DEGB gravity for fixed $\alpha= -5000\, \rm km^2$ (top panels) and $\alpha= 5000\, \rm km^2$ (bottom panels). Here the anisotropy parameter has been varied in the interval $\vert\beta\vert \leq 1.6$, where the particular case $\beta =0$ corresponds to the isotropic solutions. It is observed that the anisotropic pressure introduces notable changes on the mass-radius results at high central densities, while the variations with respect to the isotropic case are irrelevant in the low-central-density branch. For sufficiently large values of $\vert\beta\vert$ keeping negative $\beta$ with positive $\alpha$ (see blue curves in the lower panel), there exists a second stable region where $dM/d\rho_c >0$. }
    \label{FigMRRhoAniso}
\end{figure}

Figure \ref{FigMRRhoAniso} demonstrates the influence of anisotropy on WD configurations for fixed values of $\alpha$. The top panels correspond to $\alpha = -5000 \, \mathrm{km^2}$, while the bottom panels represent $\alpha = 5000 \, \mathrm{km^2}$. The anisotropy parameter $\beta$ is varied in the range $\beta \in [-1.6, 1.6]$, with $\beta = 0$ representing the isotropic case. For both values of $\alpha$ we see that the main consequence of anisotropy is an increase (decrease) in the maximum masses for positive (negative) $\beta$. However, for low central densities, the impact of anisotropy is negligible, and the results converge to those of the isotropic solutions. Therefore, anisotropic pressure plays an important role for massive WDs (i.e.~for stars with high central densities) in 4DEGB gravity. It is worth noting that this behavior is qualitatively similar to the case of NSs \cite{Pretel2020, Rahmansyah2021, Curi2022, Becerra2024} and QSs \cite{Curi2022, Pretel2024PLB, Pretel2024} in GR, where the predominant role of anisotropy only occurs at high central densities. Of course, the range of central densities and EoSs here are different from those of NSs, but the qualitative behavior regarding the impact of anisotropic pressure holds for WDs in 4DEGB gravity.

\begin{figure}[h]
    \centering
    \includegraphics[width = 5.908cm]{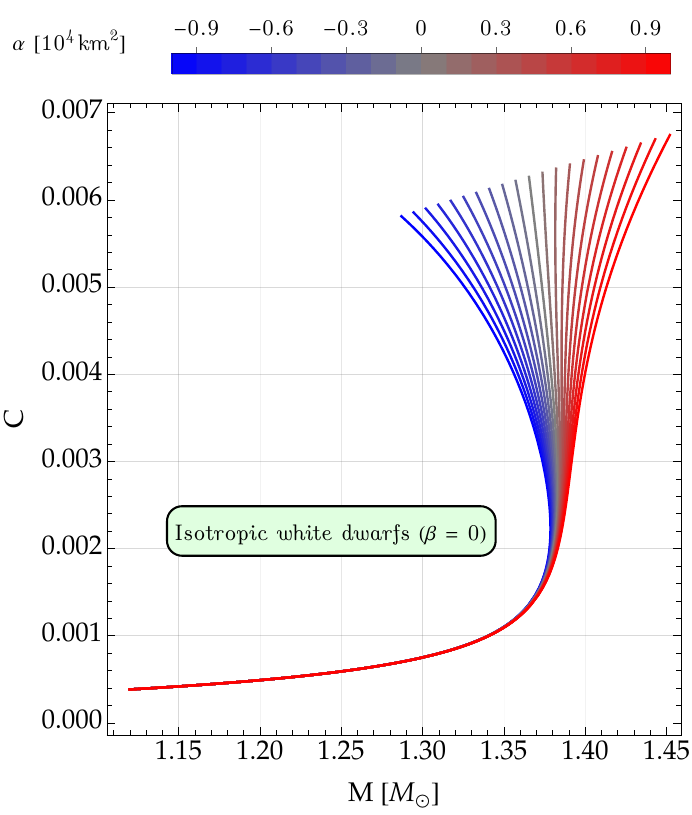}
    \includegraphics[width = 5.84cm]{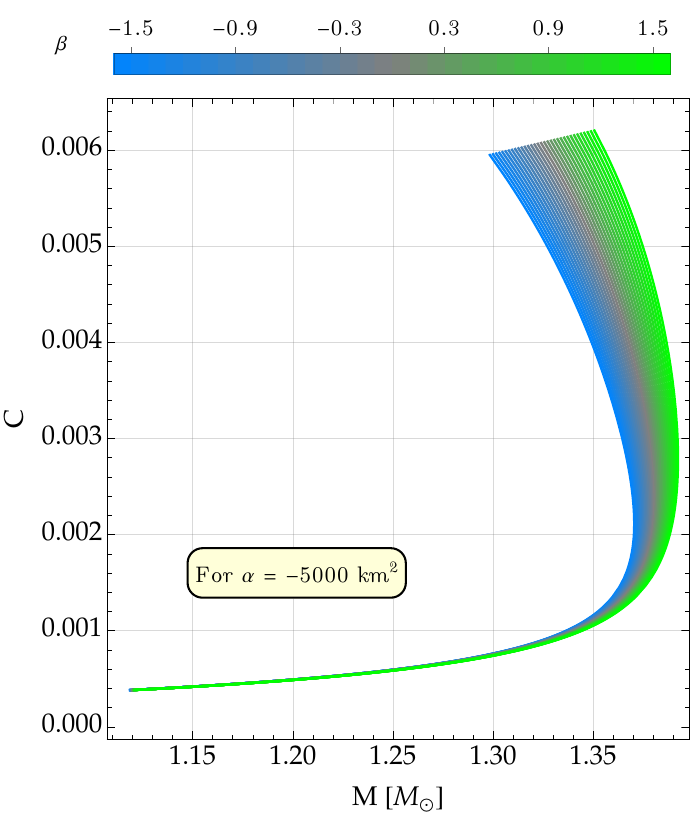}
    \includegraphics[width = 5.84cm]{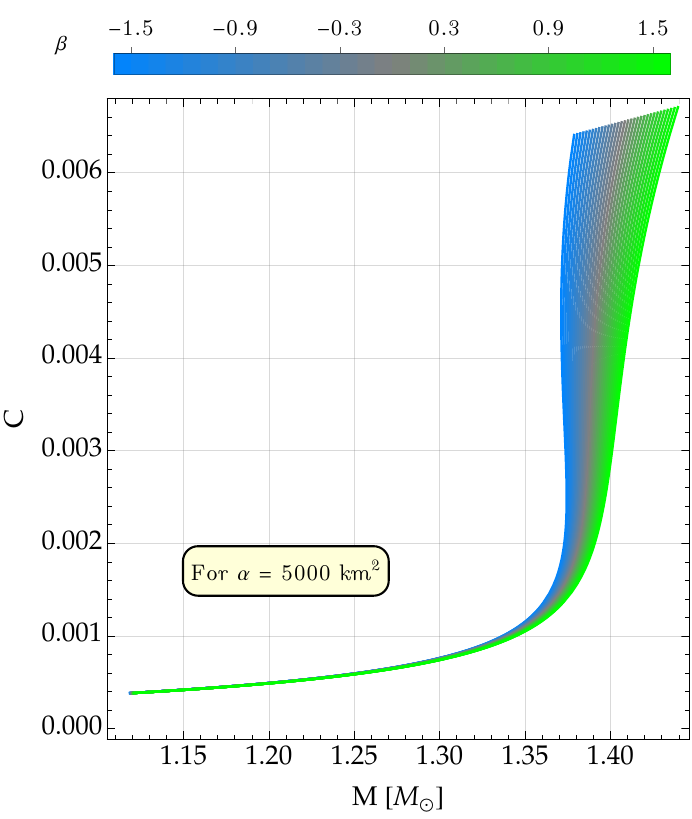}
    \caption{Compactness versus gravitational mass for the stellar configurations shown in Figs.~\ref{FigMRRho} and \ref{FigMRRhoAniso}. }
    \label{FigCompactness}
\end{figure}

The middle and right plots of Fig.~\ref{FigCompactness} show the $C-M$ relationship for $\alpha = -5000\, \rm km^2$ and $5000\, \rm km^2$, respectively. The effect of anisotropy is irrelevant for low compactnesses, but above $C \sim 0.001$ again the anisotropic pressure plays a prominent role.

\subsection{Comments on stability and observations}\label{SubsecStability}

It is well known in GR \cite{Glendenning2000, Haensel2007}, even beyond Einstein gravity \cite{Sham2012, Pretel2021JCAP, Sarmah2022}, that the turning point from stability to instability occurs when the mass is maximum, so that stable (unstable) compact stars correspond to $dM/d\rho_c >0$ ($<0$) on the $M(\rho_c)$-curves. According to the right panel of Fig.~\ref{FigMRRho}, we can obtain a maximum mass point if $\alpha \lesssim 2000\, \rm km^2$. However, above this value of $\alpha$ it is not possible to find a critical WD indicating a maximum (see the red curves). This means that the GB term (with $\alpha$ positive and sufficiently large) is capable of generating always stable massive WDs, thus overcoming the Chandrasekhar limit. This is an unprecedented and peculiar outcome of 4DEGB gravity, which is not manifested within the gravitational theory of pure GR.

In addition, the numerical results in Fig.~\ref{FigMRRhoAniso} indicate that anisotropic pressures induce noticeable changes in the mass-radius profiles, particularly at high central densities. Positive values of $\beta$ (corresponding to $p_\perp > p_r$) tend to increase the maximum mass, whereas negative values of $\beta$ have the opposite effect. Although for negative $\alpha$ there is only one stable branch, it is observed that it is possible to obtain stable anisotropic WDs on two branches in the $M(\rho_c)$-curve when $\alpha= 5000\, \rm km^2$ and for sufficiently large values of $\vert\beta\vert$ keeping negative $\beta$, see for example the curve for $\beta= -1.6$, where there are two regions on the $M(\rho_c)$-curve with positive $dM/d\rho_c$. The first critical central density $\rho_c^{\rm crit}$, where the first stability branch ceases, becomes increasingly larger as $\beta$ increases from its negative values. On the other hand, the second critical density (where $M(\rho_c)$ is a minimum) decreases with increasing anisotropy parameter $\beta$, so that the second stable branch starts at a lower $\rho_c^{\rm crit}$ value as $\beta$ grows inside the WD. From this perspective, the anisotropic pressure in WDs has a remarkable and interesting effect in the case of a positive GB coupling constant. Therefore, depending on the pair of values $\{\alpha, \beta\}$, it is possible to obtain two stable branches for anisotropic WDs in 4DEGB gravity.

The computational results underline the importance of including both isotropic and anisotropic scenarios in the analysis of WDs under 4DEGB gravity. The variation of the parameters $\alpha$ and $\beta$ leads to a diverse set of configurations, some of which could align with observed deviations from standard Chandrasekhar predictions. For instance, the presence of anisotropic pressures might explain certain anomalies in compact stellar remnants, offering a theoretical basis for their unusual properties. These results also serve as a valuable testing ground for the validity of 4DEGB gravity in astrophysical contexts.

\section{Concluding remarks and future perspectives} \label{sec:conclusion}

In this work, we have explored the relativistic structure and basic macroscopic properties of WDs within the framework of 4DEGB gravity, focusing on the impact of the GB coupling constant $\alpha$ and the effects of anisotropic pressures by means of an anisotropy parameter $\beta$. By solving the modified TOV equations numerically for a wide range of central densities, coupling constants, and pressure anisotropy parameters, we have gained valuable insights into how higher-curvature corrections influence these dense stellar remnants. 

We have shown that, when $\alpha$ is of the order $\vert\alpha\vert \leq 500\, \rm km^2$, the effects of the GB term on the $M-R$ relations of WDs are negligible. Although this range for $\alpha$ has a quite appreciable impact on the global properties of NSs \cite{Saavedra2024, Charmousis2022}, our findings here reveal an irrelevant effect for the isotropic WD case. Nevertheless, since we are dealing with another type of stellar system (i.e., a range of central densities and EoS different from those of NSs), we have also considered a larger range for $\vert\alpha\vert$ in order to appreciate notable changes in the $M-R$ diagrams of WDs. To do so, we have adopted the range $\alpha \in [-1.0, 1.0]\times 10^4\, \mathrm{km^2}$, which is within the observational constraints provided by Refs.~\cite{Clifton2020, Fernandes2022}. Consequently, our results for this larger $\vert\alpha\vert$ indicate that the GB coupling constant $\alpha$ significantly affects the equilibrium configurations of massive WDs. Positive values of $\alpha$ tend to produce smaller and more compact configurations, while negative values lead to larger and less compact structures. Importantly, the inclusion of the GB extra term allows for deviations from the Chandrasekhar mass limit, potentially providing an explanation for observed anomalies such as super-Chandrasekhar WDs \cite{Kalita2020ApJ}. The $M-R$ diagrams and $M-\rho_c$ relations obtained in this study demonstrate how $\alpha$ modifies the maximum mass and stability of WDs, where its repercussions become increasingly pronounced at high central densities.

Although the $M(\rho_c)$ method is a necessary but not sufficient condition for stability, it already provides a hint of what the radial stability of WD stars would be in 4DEGB gravity. A sufficient condition for stability would be to determine the frequencies of the adiabatic radial vibration modes. In this case, the field equations have to be linearized around the equilibrium configuration, and we hope to address this more sophisticated approach in a future study. In fact, for NSs in 4DEGB gravity, it has been shown very recently that the coincidence of the maximum-mass points with the transition to instability (where the squared fundamental eigenfrequencies are zero) still holds in this type of modified gravity \cite{Saavedra2024}. Although the stellar systems addressed in such study are different from those considered in our work, we hope that this compatibility (or coincidence) will be maintained, so that the $M(\rho_c)$ method will still hold in the case of WDs.

Furthermore, the analysis of anisotropic pressure, modeled using the QL profile \cite{Horvat2011}, further highlights the importance of including anisotropy in the study of WDs. For positive anisotropy parameters $\beta$ (where $p_\perp > p_r$), we observed an increase in the maximum mass of the star, while negative $\beta$ values reduce this global quantity. These repercussions are particularly significant at high central densities, where deviations from isotropic configurations are more pronounced. According to the classical stability criterion $dM/d\rho_c >0$ \cite{Horvat2011, Pretel2020}, our findings have shown that there exists a new branch of stable massive anisotropic WDs for some positive values of $\alpha$ and sufficiently large values of $\vert\beta\vert$ keeping negative $\beta$. In particular, for the pair of values $\{\alpha, \beta\} = \{5000\, \rm km^2, -1.6\}$, we have shown the existence of two stability regions for anisotropic WDs in 4DEGB gravity. This result already leaves an interesting avenue for future research and provides new insights into the astrophysical implications of modified gravity in dense stellar environments such as WDs. The inclusion of anisotropic pressure may also serve as a theoretical explanation for unusual stellar remnants that cannot be described by isotropic models. Indeed, the higher-curvature corrections introduced by the GB term offer a new perspective on the physics of anisotropic WDs and may help reconcile theoretical predictions with observational data. For example, our results suggest that 4DEGB gravity could provide a framework for understanding the macroscopic properties of extreme compact objects, such as those that approach or exceed the Chandrasekhar limit.

Looking ahead, there are several compelling avenues for future research based on the findings of this study. The inclusion of rotation presents a natural and essential extension, as rapid rotation introduces significant centrifugal support that can alter the equilibrium configurations of WDs. Investigating the interplay between rotation, the GB coupling, and anisotropic pressures would provide a more comprehensive understanding of these compact objects. Additionally, the stability of WDs under perturbations remains a critical aspect that warrants further exploration. Studying radial and non-radial oscillation modes in the context of 4DEGB gravity could offer new insights into stability criteria and help constrain the GB coupling constant. Another key direction involves comparing theoretical predictions with observed properties of WDs, particularly those in binary systems or acting as progenitors of Type Ia supernovae. Such comparisons, especially for systems that exceed the Chandrasekhar limit in modified gravity scenarios, could serve as vital tests of 4DEGB gravity. Lastly, the strong magnetic fields often observed in WDs present another layer of complexity, as they may interact with GB corrections in non-trivial ways. Incorporating magnetic-field effects into the modeling would enhance the realism and applicability of the results, providing a richer understanding of the dynamics and structure of these fascinating stellar remnants. \textcolor{black}{It is worth emphasizing that this gravity theory needs to be examined within the context of solar-system tests, as was done for example in $f(R)$ gravity \cite{Guo2014, Negrelli2020}, and we will also leave this for a future study.}

In conclusion, the results presented here contribute to the growing body of research on modified gravity and its implications for astrophysics. WDs provide a robust platform for testing the consequences of higher-curvature corrections, and future studies that incorporate additional physical effects and observational constraints will further enhance our understanding of these fascinating objects. By bridging the gap between theoretical predictions and astronomical observations, 4DEGB gravity holds the potential to reshape our understanding of compact stars and their role in the cosmos.

\section*{date availability}

There are no new data associated with this article.

\begin{acknowledgments}
We gratefully acknowledge the insightful comments and constructive suggestions provided by the Editor and the anonymous referee. Their valuable feedback has significantly improved both the clarity and the physical grounding of our manuscript. JMZP acknowledges support from ``Funda\c{c}{\~a}o Carlos Chagas Filho de Amparo {\`a} Pesquisa do Estado do Rio de Janeiro'' -- FAPERJ, Process SEI-260003/000308/2024. T. T. was supported by Walailak University under the New Researcher Development scheme (Contract Number WU67268). He also acknowledges COST actions CA21106 and CA22113. \.{I}.~S. expresses gratitude to EMU, T\"{U}B\.{I}TAK, ANKOS, and SCOAP3 for their academic and/or financial support. He also acknowledges COST Actions CA22113, CA21106, and CA23130 for their contributions to networking.
\end{acknowledgments}\

\end{document}